\documentstyle[seceq,epsf,wrapft]{ptptex}
\notypesetlogo  %comment in if to eliminate PTPTeX logo
\title{Clustering and Correlations in Neutron Haloes}

\author{
Nigel {\sc Orr}\footnote{e-mail: orr@caelav.in2p3.fr} 
}

\inst{LPC-ISMRa, Bd Mar\'echal Juin, 14050 Caen Cedex, France
}

\recdate{%      %Editorial Office will fill in this.
\today
}

\abst{In the present paper clustering and correlations 
within halo systems
is explored.  In particular, the application of neutron-neutron
interferometry and Dalitz-plot type analyses is presented through
the example provided by the dissociation of $^{14}$Be.  
A novel approach for producing and detecting bound neutron clusters
is also described.  The observation of some 6 events with characteristics
consistent with the liberation of a multineutron cluster in the breakup of 
$^{14}$Be -- possibly in the channel $^{10}$Be+$^4$n -- is discussed.
}

\begin{document}

\maketitle

\section{Introduction}

Clustering, which has long been known to occur along the line
of beta stability,
also appears in more exotic forms as the drip-lines are approached. 
For example, $2\alpha-Xn$ molecular-like configurations have been observed in
excited states of $^{10,12}$Be\cite{Fre99}.
The most spatially extreme form of clustering are the neutron haloes
which occur as the ground states of some nuclei at the limits of particle 
stability. Perhaps the most intriguing 
of the halo systems are the Borromean two-neutron halo nuclei ($^{6}$He, 
$^{11}$Li and 
$^{14}$Be), in which the two-body subsystems (core-$n$ and $n-n$) are 
unbound. Such 
behaviour naturally gives rise to the question of the correlations between the 
constituents. Even in the case of the most studied 
of these nuclei, $^{6}$He and $^{11}$Li, little is known in this respect.
In the first part of this paper we explore the nature of these correlations
through the application of interferometry and Dalitz-plot type analyses to 
kinematically complete measurements of dissociation.

On a more speculative note, the production and detection of bound multineutron 
clusters in the breakup
of very neutron-rich secondary beams is explored in the final section of this paper.  
This approach exploits the possibility that
multineutron halo nuclei and other very neutron-rich systems contain components
of the wavefunction in which the neutrons exist in a relatively compact cluster-like
configuration. 
A new method is introduced here for the direct detection of neutral clusters and the results
obtained from an analysis of data acquired with beams of $^{11}$Li and $^{14}$Be
is presented.  

\section{Experimental Setup}

The results described here were derived from data taken using what is now a
relatively standard experimental configuration for kinematically complete
measurements of the breakup of neutron-rich beams\cite{Lab01}.
The secondary beams ($\overline{E}$=30-50~MeV/nucleon) were prepared from a
63 MeV/nucleon $^{18}$O primary beam using the LISE3 spectrometer at GANIL.  
The beam particles
were tracked onto the breakup targets (C and Pb) 
using two
position sensitive parallel plate avalanche counters.  The charged fragments
from dissociation were identified using a large area position
sensitive Si-CsI telescope 
centred at zero degrees and located $\sim$15~cm downstream of the target.  

The neutrons emitted at forward angles were detected using the 99 elements of the
DEMON array.  The array covered angles between +13$^\circ$ and
$-$40$^\circ$ in the horizontal plane and $\pm$14$^\circ$ in the  vertical with the
modules arranged in a staggered configuration at distances between 2.5 and 6.5 m 
from the target\cite{Mar00}.  Such a geometry provided for a relatively high
two-neutron detection efficiency (1.5\%) whilst reducing the rate of cross-talk ---
both intrinsically and via the use of an off-line rejection algorithm --- to
negligible levels\cite{Mar00,Lab01}.  

\section{Correlations in Two-Neutron Halo Nuclei}

We have explored the spatial configuration of the halo
neutrons at breakup through the 
application of the technique of intensity interferometry --
an approach first developed for stellar interferometry by Hanbury-Brown and Twiss in 
Australia in the
1950's and 60's\cite{HBT} and later extended to source size measurements in high energy
collisions\cite{Gol60}.  The principle behind the technique is as follows:  
when identical particles are emitted 
in close proximity in space-time, the wave function
of relative motion is modified by the FSI and quantum statistical
symmetries\cite{Boa90} --- in the case of halo neutrons the overwhelming effect is that of 
the FSI\cite{FMM00}. Intensity interferometry relates this
modification to the space-time separation of the particles at emission as a 
function of the four-momenta of the particles through the correlation function 
$C_{\rm{nn}}$, which is defined as,

\begin{equation}
 C_{\rm{nn}}(p_1,p_2)=\frac{d^2n/dp_1dp_2}{(dn/dp_1)\,(dn/dp_2)} \label{e:C12}
\end{equation}

where the numerator is the measured two-particle distribution and the denominator the 
product of the independent single-particle distributions\cite{FMM00}.  
As is generally the case, the single-particle distributions
have been generated in our work via event mixing.  Importantly, in the 
case of halo neutrons
special consideration must be given to the strong residual correlations\cite{FMM00}.
Experimentally care needs to be taken to eliminate
cross talk\cite{Mar00}.

As a first step, the measurements of breakup on a Pb target
of $^{6}$He, $^{11}$Li and 
$^{14}$Be were analysed\cite{FMM00}.  The choice of a high-Z target was made to 
privilege
Coulomb induced breakup, whereby the halo neutrons may in a first approximation 
act as spectators and for which simultaneous emission may be 
expected to occur.  The correlation functions derived from the data, assuming 
simultaneous emission, were compared to
an analytical formalism based on a Gaussian source\cite{soviet}.
Neutron-neutron separations of $r_{nn}^{RMS}=5.9\pm1.2$~fm ($^{6}$He), 
$6.6\pm1.5$~fm ($^{11}$Li) and $5.6\pm1.0$~fm ($^{14}$Be) were thus extracted.  These 
results appear to
preclude any strong dineutron component in the halo wavefunctions at breakup; a
result which, for $^{6}$He is in line with a recent radiative capture experiment we
have performed\cite{Sau01}.
It is interesting in this context to compare these results to the RMS neutron-proton
separation of 3.8~fm in the deuteron (the only bound two nucleon system).

The same analysis has been applied to dissociation of $^{14}$Be by a C target,
in order to investigate the influence of the reaction mechanism.  A 
result which hints at a somewhat larger separation, $r_{nn}^{rms}=7.6\pm1.7$~fm, was 
obtained. This raises the question as to whether simultaneous emission can be
assumed a priori. In principle, the analysis of the correlation function in two
dimensions, transverse and parallel to the total momentum of the pair, would
allow for the unfolding of the source size and lifetime\cite{soviet}. Such an
analysis requires a much larger data set than presently available. 
The two-neutron halo, however, is far less complex than
the systems usually studied via interferometry (for example, heavy-ion 
collisions\cite{Boa90}).
Moreover, the simple three-body nature of the system breaking up suggests
that any delay in the emission of one of the neutrons will arise
from core-n FSI/resonances in the exit channel, a process that may be expected to be
enhanced for nuclear induced breakup.

\begin{figure}
\epsfxsize=9.5cm
\centerline{\epsfbox{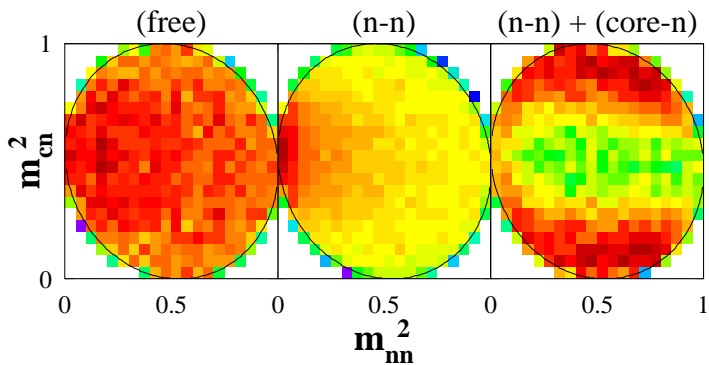}}
\caption{Dalitz plot for the simulated decay of $^{14}$Be (see text).  
In the left panel no FSI are included.}
\end{figure}

Correlations in three-particle decays are commonly encountered in particle
physics and are typically analysed using plots of the
squared invariant masses of particle pairs $(M_{ij}^2,M_{ik}^2)$, with
$M_{ij}^2=(p_i+p_j)^2$; a technique developed by the Australian physicist Richard Dalitz
in the early 1950's\cite{Dal53}. In Dalitz-plot representations, 
FSI or resonances lead to a
non-uniform population of the surface within the kinematic boundary defined by
energy-momentum conservation and the decay energy. 
In the present case, the core+n+n system exhibits a distribution of decay
energies ($E_{\rm{d}}$). The $E_{\rm{d}}$ associated 
with each event
will thus lead to a different kinematic boundary, and the resulting plot
containing all events cannot be easily interpreted. We have thus introduced a
normalised invariant mass, 

\begin{equation}
 m_{ij}^2 = \frac{M_{ij}^2-(m_i+m_j)^2}{(m_i+m_j+E_{\rm{d}})^2-(m_i+m_j)^2}
\end{equation}

which ranges between 0 and 1 
(that is, a relative energy $E_{ij}=M_{ij}-m_i-m_j$ 
between 
0 and $E_{\rm{d}}$) 
for all events and exhibits a single kinematic boundary. 
Examples of how n-n and core-n FSI may manifest themselves 
in the Dalitz plot for the decay of $^{14}$Be are illustrated in Fig.~1, whereby 
events have been simulated according to the simple interacting phase-space
model described in ref.\cite{FMM01}. The inputs were an 
$E_{\rm{d}}$ distribution following that measured\cite{Lab01}, the 
$C_{\rm{nn}}$ obtained with the C target, and a
core-n resonances with $\Gamma=0.3$~MeV at $E_0=0.8$~MeV.  Note that due to the
normalisation the (squared) core-neutron invariant mass does not present 
a simple structure directly
related to the energy of the resonance/FSI\cite{FMM01}.

\begin{figure}
\epsfxsize=8cm
\centerline{\epsfbox{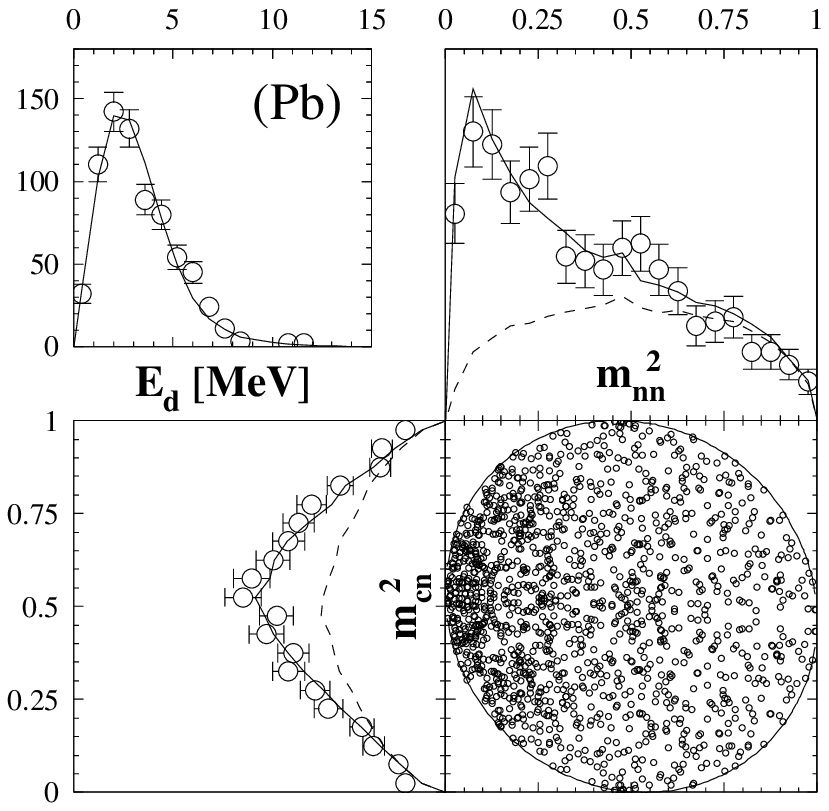}}
\epsfxsize=8cm
\centerline{\epsfbox{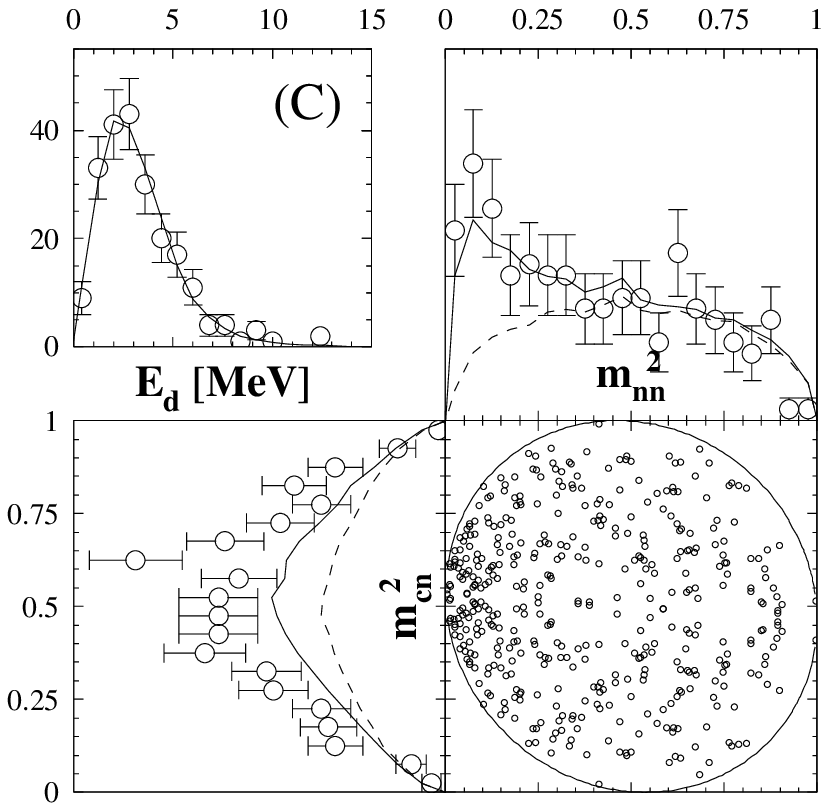}}
\caption{Dalitz plot and the projections onto the squared invariant masses
for the dissociation of $^{14}$Be by Pb (upper) and by C (lower panels). 
The lines are the 
phase-space model simulations
with/without (solid/dashed) n-n FSI. The inset shows the measured $E_d$ spectrum.}
\end{figure}

The Dalitz plot for the data from the dissociation by Pb (Fig.~2, upper panel)
presents a strong n-n FSI and a uniform density for 
$m_{\rm{nn}}^2$>$\sim0.5$.
Indeed, the n-n FSI alone describes very well the projections onto both axes,
and therefore suggests that core-n resonances are not present to any
significant extent. This result confirms the hypothesis of simultaneous n-n
emission employed in the original analysis of the dissociation of $^{14}$Be by 
Pb\cite{FMM00}. The $r_{nn}^{RMS}$ so extracted, $5.6\pm1.0$~fm, may thus
be considered to represent
the n-n separation in the halo of $^{14}$Be.
 
For dissociation by the C target (Fig.~2, lower panel), despite the lower
statistics, two differences are evident. Firstly, the n-n signal is weaker,
indicating that a significant delay has occurred between
the emission of each neutron. Second, and more importantly, the agreement
between the model including only the n-n FSI and the data for $m_{\rm{cn}}^2$
is rather poor. In order to verify whether this disagreement corresponds to the
presence of core-n resonances the core-n relative energy, $E_{\rm{cn}}$, has
been explored.
It has been reconstructed for the simulations incorporating only the n-n FSI
and compared in Fig.~3 to the data (the model calculations have been
normalized to the data above 4~MeV). For dissociation by Pb, the inclusion of
only the n-n FSI provides a very good description of the data, with the
exception of small deviations below 1~MeV. This is in line with the Dalitz-plot 
analysis discussed above.

The deviations observed for the C target between the measured $m_{\rm{cn}}^2$
and the simulation including only the n-n FSI clearly
correspond to structures in the $E_{\rm{cn}}$ spectrum. Moreover, these
structures are located at energies that are in line with those of states
previously reported in $^{13}$Be: the supposed $d_{5/2}$ resonance at 
2.0~MeV\cite{Ost92,Bel98} and a lower-lying state(s)\cite{Orr00,Bel98,Tho98}.
The model-to-data ratio is about 1/2, indicating that the peaks correspond to
resonances formed by one of the neutrons in almost all decays; the solid line 
accounts for the contribution of the neutron not 
interacting with the core. If we add to the phase-space model with n-n FSI
core-n resonances ($\Gamma=0.3$~MeV) at $E_0=0.8$, 2.0\cite{Bel98} and 
3.5~MeV\footnote{The present data are not particularly
sensitive to the location and form of the states, in particular below 1~MeV, 
and a level at 0.5~MeV would, for example, equally well describe the data.} 
with intensities of 45, 35 and 20\%, respectively, the data are well reproduced 
(dashed line). In the case of dissociation by Pb, the lowest-lying level(s) 
appears to be present in at most 10\% of events.
 
In the context of the influence of the reaction mechanism, it is worthwhile 
noting that whilst some 35\% of the two-neutron removal 
cross section on the Pb target is  
attributable to nuclear 
induced breakup\cite{Lab01}, the requirement of two neutrons in coincidence 
with the
$^{12}$Be core in the present analysis reduces 
this to some 15\% -- approximately half
of the two-neutron removal cross section arises from absorption.  

By combining the information extracted from the core-n channel with the n-n
correlation functions, the analysis can be extended to extract the average
lifetime of the core-n resonances. If the n-n separation in $^{14}$Be is fixed to
that obtained for dissociation by Pb, $r_{nn}^{RMS}=5.6\pm1.0$~fm, the 
delay between the emission of the neutrons $\tau_{\rm{nn}}$ needed to describe 
the n-n correlation function for the C target may be introduced. As discussed above, 
this delay
should correspond to the lifetime of the resonances. The 
result of a $\chi^2$ analysis, represented by the dashed lines in 
Fig.~3 (bottom right panel), suggests an average lifetime of $150^{+250}_{-150}$~fm/$c$.

\begin{figure}
\epsfxsize=10cm
\centerline{\epsfbox{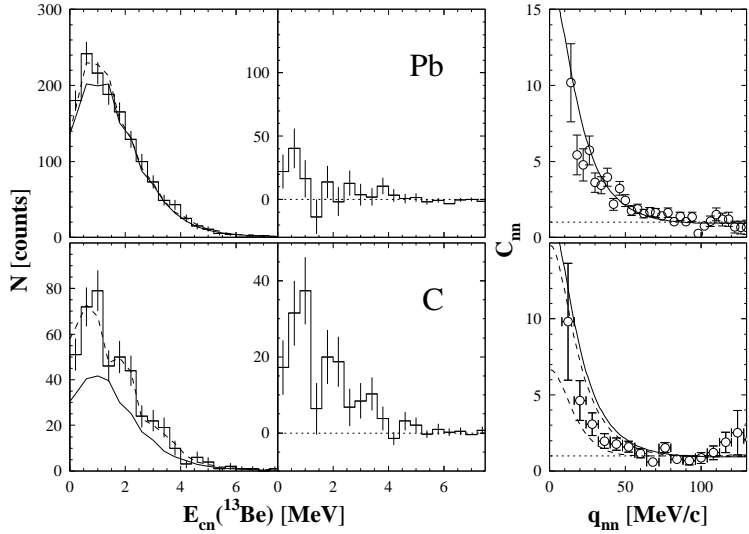}}
\caption{Core-n relative energy distributions (left) and n-n correlation 
functions (rightmost panels) for the dissociation of $^{14}$Be by Pb and C. The 
lines in the $E_{\rm{cn}}$ spectra are the result of the phase-space model
simulations with n-n FSI (solid) plus core-n FSI (dashed, see text). The
histograms presented in the middle panels are the difference between the data 
and the n-n FSI simulations. The solid lines in the panels at the right are the 
$C_{\rm{nn}}$ for $r_{nn}^{RMS}=5.6$~fm and $\tau_{nn}=0$; the dashed lines 
correspond to the limits of the range $r_{nn}^{RMS}=6.6$--4.6~fm  and 
$\tau_{\rm{nn}}=0$--400~fm/$c$.} 
\end{figure}

\section{Multineutron Clusters}

The very lightest nuclei have long played a fundamental role in testing nuclear
models and the underlying nucleon-nucleon interaction. 
In this
context the study of systems exhibiting very asymmetric $N/Z$ ratios may
provide new perspectives on the nucleon-nucleon interaction and few-body forces. In the
case of the light, two-neutron halo nuclei such as $^6$He, insight is already
being gained into the effects of the three-body force\cite{Zhu93}. Very
recently evidence has been presented that the ground state of $^5$H exists as a
relatively narrow, low-lying resonance\cite{Kor01}. In the case of the
lightest $N=4$ isotone, $^4$n, nothing is known\cite{Til92,Ogl89}. The
discovery of such neutral systems as bound states would have far reaching
implications for many facets of nuclear physics. 

It is, therefore, interesting to speculate that multineutron halo nuclei
and other very neutron-rich systems may contain components 
of the wavefunction in which
the neutrons present a relatively compact cluster-like configuration.  If this
were to be the case, then the dissociation of beams of such nuclei may offer a means
to produce bound neutron clusters (if they exist) and, more generally
study multineutron correlations.  

To date the majority of searches for 
multineutron systems have relied on very low (typically $\sim$1~nb) cross 
section double-pion 
charge exchange (D$\pi$CX) and heavy-ion transfer
reactions (see, for example, refs\cite{pion,HI}).  In the case of dissociation of
an energetic beam of a very neutron-rich nucleus, relatively high cross sections
(typically $\sim$100~mb) are encountered.  Thus, even only a small component
of the wavefunction corresponding to a multineutron cluster could result 
in a measurable
yield with a moderate secondary beam intensity.  Furthermore the backgrounds
arising in D$\pi$CX and heavy-ion transfer reactions from target impurities 
and complex
many-body phase space
reactions are obviated in breakup.

The difficulty in this approach lies in the direct detection of a $^A$n cluster.
The avenue that we have explored is to detect the recoiling
proton in a liquid scintillator\cite{FMM01a}.  One of the principle advantages 
of a liquid scintallator is that neutrons may be discriminated with good efficiency from
the $\gamma$ and cosmic-ray backgrounds using pulse-shape analysis. 
Careful source and cosmic-ray calibrations\footnote{The astute reader will also
realise that the maximum proton recoil energy for a given $E_n$ may also be used
for calibration purposes -- a procedure that we are currently employing.} 
permit the charge deposited and hence the energy ($E_p$) of the recoiling proton  
to be determined.
This may be compared to the energy derived from the measured time-of-flight
($E_n$):
for a single neutron and an ideal detector, $E_p$/$E_n$$\leq$1; for a 
realistic detector with 
finite resolution the limit is $\sim$1.4.  In the case of a 
multineutron cluster ($^A$n) $E_p$ can exceed the incident energy per nucleon
and
$E_p$/$E_n$ will take on a range of values 
extending beyond 1.4 --- up to $\sim$3 in the case of A=4 (Fig. 4).

\begin{figure}
\epsfxsize=7cm
\centerline{\epsfbox{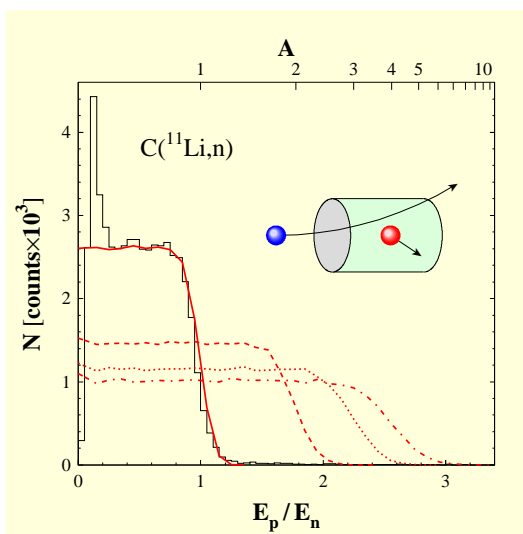}}
\caption{$E_p/E_n$ for A=1 (solid line), 2 (dashed), 3 (dotted) and 4 (dot-dashed).
In the case of A=1, comparison is made to single neutron events from the $^{11}$Li
breakup of $^{11}$Li. The excess of events at low $E_p/E_n$ arise from reactions on
the carbon component of the scintillator.}
\end{figure}  

As discussed in ref.\cite{FMM01a}, the DEMON
modules exhibit saturation effects at very high light output\footnote{The initial 
goal
of the experiment run to acquire the data analysed here was not the search for
multineutron clusters, and as such the analysis of very high light outputs was
not foreseen.  The possibility of operating the DEMON photomultipliers
at lower voltages is being explored for future dedicated experiments.}.
In order to avoid this problem,
particularly in the region $E_p/E_n$>1, an upper limit of
$E_{\rm{n}}=18$~MeV/nucleon
was imposed.

At low energies the proton recoil is free of saturation effects. However,
background events arising from $\gamma$ and cosmic rays represent a potential
contaminant for the $E_p/E_n$ distribution. These events are randomly distributed
in time, and thus the relative rate increases at low energy 
since $E{\propto}t^{-2}$. As the energy loss in a module is completely
uncorrelated with the inferred time-of-flight,
$E_p/E_n$ is not confined to values inferior to 1.4. Even 
if the rejection rate using pulse-shape
analysis is close to 100~\%, any events that remain could mimic a $^A$n signal.
As detailed in ref.\cite{FMM01a}, a lower limit on
$E_{\rm{n}}$ of 11~MeV/nucleon was thus imposed.  Importantly, any such events which
do survive these conditions should not be correlated with any particular breakup
fragment.

\begin{figure}
\epsfxsize=8.5cm
\centerline{\epsfbox{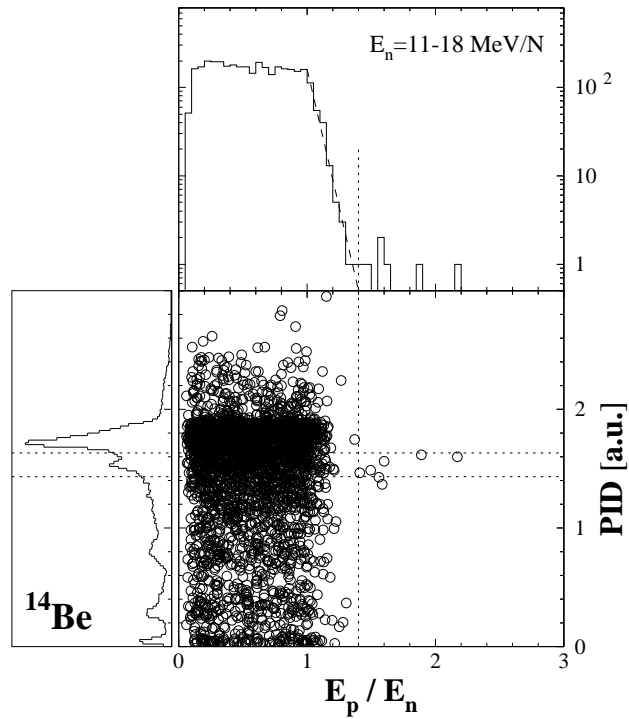}}
\caption{PID versus $E_p/E_n$ for the 
reaction ($^{14}$Be,X+n).  The prominent peak at PID$\sim$1.7 corresponds to
$^{12}$Be fragments.
The horizontal
band (dotted line) corresponds to the range of PID values encompassing the $^{10}$Be
fragments. }
\end{figure}

\begin{figure}
\epsfxsize=8.5cm
\centerline{\epsfbox{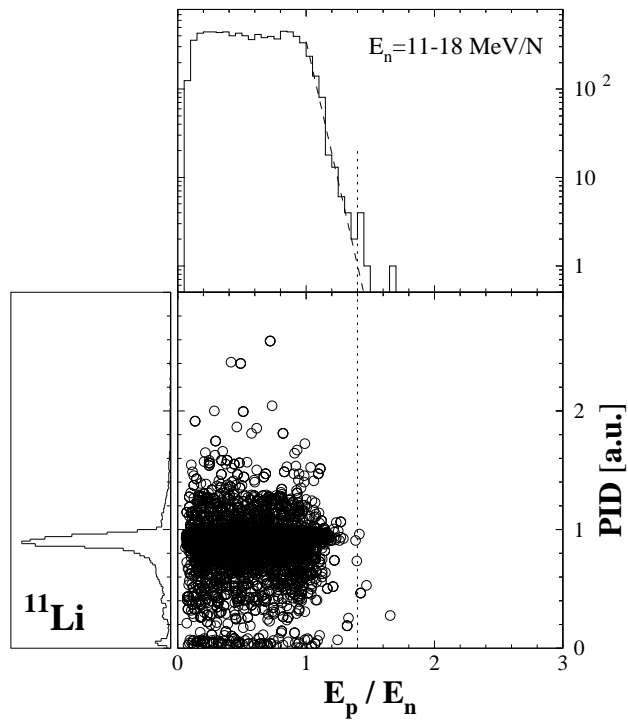}}
\caption{PID versus $E_p/E_n$ for the 
breakup of $^{11}$Li.  
The prominent peak at PID$\sim$9 corresponds to $^{9}$Li fragments.}
\end{figure}  

The data already at hand from the study of the disociation of $^{14}$Be and 
$^{11}$Li\cite{Lab01,FMM00,FMM01}
was examined with a view to testing the method outlined above.
The details of the analyses carried out may be found in ref.\cite{FMM01a}.  The 
essential results are provided by figures 5 and 6 which display the charged
fragment particle identification (PID) derived from the Si-CsI detector 
telescope versus $E_p/E_n$.  

The $E_p/E_n$ distributions (upper panels in Figs. 5 and 6)
exhibit a general trend below 1.4: a plateau up to 1 followed by a sharp
decline, which may be fitted to an exponential distribution (dashed line). In
the region where $^A$n events may be expected to appear some 7
events with $E_p/E_n$ ranging from 1.4 to 2.2 are observed for $^{14}$Be. 
In the case
of $^{11}$Li, despite the greater number of neutrons detected (factor of 2.4),
only 4 events appear which lie just above threshold. Turning to the 
coincidences with the charged fragments, the 7 events produced by the $^{14}$Be beam fall
within a region centred on $^{10}$Be. In the case of the 4 events produced in
the reactions with $^{11}$Li no correlation appears to exist with 
any particular fragment.

The left panel in figure 7 displays in more detail the region of the
particle identification spectrum for the breakup of $^{14}$Be into lighter Be
isotopes, together with the 7 events in question. Clearly the
resolution in PID does not allow the
observed events to be unambiuously associated with a $^{10}$Be fragment. However, 
the much higher
cross-section for this channel (460$\pm$40~mb) compared to $^{11}$Be (145$\pm$20~mb)
suggests that this may be the case.
It should be noted that the PID is somewhat complicated by the fact that reactions also occur in the
Si-CsI telescope.  The effects of this are shown in figure 7 (right panel), whereby
the reactions in the telescope give rise to a tail extending to higher 
mass Be fragments.
Ideally a dedicated experiment including a high statistics 
target-out measurement
would remove this ambiguity.

\begin{figure}
\epsfxsize=8cm
\centerline{\epsfbox{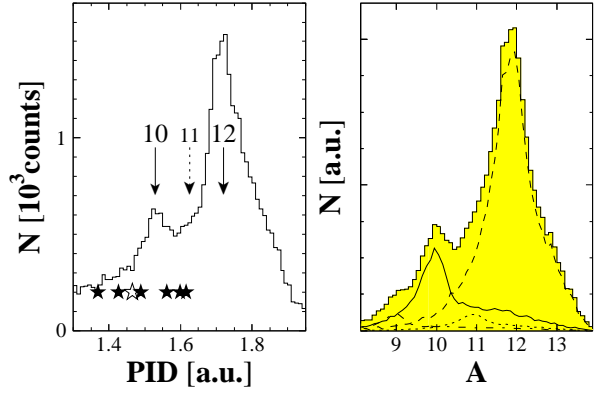}}
\caption{Left: detail of the particle identification spectrum around
$^{10,12}$Be for the data from the reaction ($^{14}$Be,X+n); the 7 events with
$E_p/E_n>1.4$ are denoted by the symbols. Right: results of a simulation
of the reactions ($^{14}$Be,$^{9-12}$Be) in the target and telescope; the shaded
histogram is the sum of the contributions from all four fragments.}
\end{figure}

As a first step towards investigating the nature of the events with $E_p/E_n$>1.4 
each was examined to verify that it corresponded to a well
defined event in both the charged
particle and neutron detectors.  Of the 7 events observed in the breakup of $^{14}$Be,
all but one survived.  
The 6 remaining events thus appear to exhibit 
characteristics consistent with detection
of a multineutron cluster from the breakup of $^{14}$Be.  Potential sources
of such events not involving the formation of a multineutron were consequently
examined\cite{FMM01a}.  
  
\begin{table}
\begin{center}
 \caption{Comparison of the number of events observed (exp) with $E_p/E_n>1.4$ for each
channel with the estimated number of events expected
from pile-up. The methods are based on a Monte-Carlo simulation (sim),
and the relative-angle distribution of n-n pairs (nn). 
The latter is quoted in terms of a conservative upper limit\cite{FMM01a}.}
 \begin{tabular}{lllll} \noalign{\medskip}\hline\hline\noalign{\smallskip}
 Channel & $N_{\rm{2n}}^{\rm{exp}}$ & $N_{\rm{2n}}^{\rm{(sim)}}$ &
 $N_{\rm{2n}}^{\rm{(nn)}}$ \\ \noalign{\smallskip}\hline\noalign{\smallskip}
 ($^{11}$Li,X)         & 4  & $\sim$3 & $<$7.0 \\
 ($^{14}$Be,$^{12}$Be) & 0  &       0.8 & $<$1.2 \\
 ($^{14}$Be,$^{10}$Be) & 6  &       0.2 & $<$0.8 \\
 \noalign{\smallskip}\hline\hline
 \end{tabular}
\end{center}
\end{table}

The most obvious source of events that may mimic a multineutron cluster is the 
detection, in the same event, of more than one neutron in the same module.
The rates at which such pile-up is expected to occur have been examined in detail
employing both simulations which reproduce the observed neutron 
angular, energy and multiplicity distributions, together with an analysis based on the
measured neutron-neutron relative angle distributions\cite{FMM01a}.
As summarised in Table~I, the two methods provide consistent results which are in 
line with
the numbers of events observed for the channels ($^{11}$Li,X+n) and ($^{14}$Be,$^{12}$Be+n).
In the case of ($^{14}$Be,$^{10}$Be), less than one event arising from pile-up is estimated
to occur with $E_p/E_n$>1.4, compared to some 6 observed events.
Given such results we conclude that reasonable evidence exists for the 
production of a multineutron cluster in the breakup of $^{14}$Be --- most probably in the
channel $^{10}$Be+$^4$n.

The average flight time of the 6 events from the target to DEMON is
$\sim100$~ns. This indicates that the lifetime must be of this order or longer.
The conditions applied in the analysis make an estimate of the
production cross-section rather difficult. Nonetheless, if we
assume that these conditions affect the 
number of neutrons and $^4$n
in a similar manner, we can scale the cross-section measured for the production
of $^{10}$Be\cite{Lab01} by the relative yield observed and obtain
$\sigma(\mbox{$^4$n})\sim1$~mb.

\section{Conclusions}

An experimental programme to explore clustering and correlations in
halo systems has been described.  New approaches have been
developed, including the application of neutron-neutron interferometry and Dalitz-plot
analyses to the dissociation of two-neutron halo nuclei.  Attempts to produce and detect
directly bound multineutron clusters have also been described and the results from a
measurement of the breakup of $^{14}$Be discussed.

Very recently a high statistics measurement of the dissociation of $^{6}$He has been carried
out.  Given that $^{6}$He is structurally the most well known two-neutron halo system, this
work should provide a good test of the techniques described here to probe correlations.   
Furthermore, correlation function analyses employing the longitudinal and transverse
neutron-neutron relative momenta should provide an independent means to disentangle the
halo neutron-neutron separation and time delay in emission.  Measurements in the 
coming year with
a $^{8}$He beam should allow multineutron correlations to be explored.

In terms of neutron clusters the confirmation or otherwise of the events 
observed here with a higher intensity $^{14}$Be
beam and improved fragment detection system is planned for the coming year.  
The search for
similar events in the breakup of $^8$He will also be undertaken. 
The saturation effects encountered with DEMON
at high light outputs will be reduced by lowering the beam
energy (in the case of the $^8$He measurement) as well as the 
high voltage applied to the 
photomultipliers. In the longer
term, searches for heavier multineutron clusters could be envisaged when more
neutron-rich beams become available at intensities beyond $10^2$~pps.

\section{References}

\section*{Acknowledgements}

I would
like to draw special attention to the key r\^oles played by
Miguel Marqu\'es (correlations and neutron clusters) 
and Marc Labiche ($^{14}$Be breakup) in the work described here.
It is also a pleasure to thank the members of the E295 collaboration and, in
particular,  
the DEMON and CHARISSA crews for their contributions. 
The support provided by the staffs of LPC (in particular JM Gautier, 
P Desrues, JM Fontbonne, 
L Hay, D Etasse, J Tillier) and GANIL (R Hue, C Cauvin, R Alves Conde) 
in preparing and executing
the experiments is gratefully acknowledged.   

This work was funded under the
auspices of the IN2P3-CNRS (France) and EPSRC (United Kingdom). Additional
support from the ALLIANCE programme (Minist\`ere des Affaires Etrang\`eres and
British Council) and the Human Capital and Mobility Programme of the European
Community (Access to Large Scale Facilities) is also acknowledged.

\end{document}